\documentclass[12pt]{article}
\title{Enhanced Electroweak Radiative Corrections in SUSY: Gluon-free
Observables}
\author{I.V.Gaidaenko, A.V.Novikov, V.A.Novikov \\
ITEP, Moscow, Russia \\
A.N.Rozanov \\
ITEP and CPPM, IN2P3, CNRS, Marseille, France \\
M.I.Vysotsky \\
INFN, Sezione di Ferrara, Italy and ITEP}
\date{}
\begin{document}
\maketitle

\begin{abstract}

Large top quark mass is responsible for the enhancement of the
oblique radiative corrections in SUSY models. We present the
analytical formulas for these corrections to the $W$-boson mass $m_W$
and to $Z$ $l^+ l^-$ coupling constants. The comparison with the result
of the Standard Model fit is made.

\end{abstract}

\newpage

The precision measurements of the $Z$-boson parameters, $W$-boson
mass and top-quark mass have demonstrated that a Standard Model
perfectly describes this part of Physics \cite{1}. One has no hopes
to improve a Standard Model fit by introducing New Physics since it
is already perfect. For the data set announced at HEP97 Conference we
obtained $\chi^2/d.o.f. = 18/14$ \cite{2}.

Nevertheless, the people still believe that there should be physics
beyond the Standard Model and the most popular candidate is
supersymmetry. In SUSY models there is a natural reason for the
success of the Standard Model in describing the data at the energy
scale of the order of the intermediate boson masses. This reason is
decoupling: the contribution of SUSY particles to the "low energy"
observables is suppressed like $(m_{W,Z}/m_{SUSY})^2$. That is why,
when comparing the results of the SUSY model calculations with the
experimental data, one gets the lower bounds on superpartner masses.
To perform such a program one-loop electroweak radiative corrections
in SUSY models should be calculated and this was done, see e.g. 
 \cite{3} -
\cite{5}. However, since there are a lot of one-loop diagrams and a
large number of parameters even in the simplest MSSM the qualitative
picture of electroweak radiative corrections did not emerge. The aim
of our investigation is two-fold. Firstly, we will present  simple
analytical formulas which describe the main part of the radiative
corrections in a wide class of SUSY models; secondly, we will include
SUSY particles in the description of the electroweak radiative
corrections developed and reviewed in \cite{6}.

In the present paper we will deal with the observables which are less
sensitive to the strong interactions, i.e. with the mass of the
$W$-boson and with $Z$ $l^+ l^-$-coupling constants $g_A$ and $g_V$.
In future we plan to incorporate $Z$-boson decays to hadrons which
will enable us to perform the general fit of the data and to get lower
bounds on the masses of Superpartners.

In order to calculate the radiative corrections to $m_W$, $g_A$ and
$g_V$ one starts with the corrections to $G_{\mu}$, $m_Z$ and
$\bar{\alpha}\equiv \alpha(m_Z)$. The corrections to $G_{\mu}$ are
described in SUSY theories by box and vertex diagrams and by the
correction to $W$-boson propagator. The corrections to $m_Z$ and
$\bar{\alpha}$ are described by self-energy insertions. Having
calculated all the terms one comes to the corrections to $m_W$, $g_A$
and $g_V$. The corrections to $m_W$ contain self energy insertions
and those to the coupling constants contain the vertices as well. In
general all the above corrections are of the order of
$(m_{W,Z}/m_{SUSY})^2$ and there are a lot of contributions from a large
number of diagrams. However, the violation of $SU(2)_V$ symmetry by
large top quark mass penetrates into the SUSY sector of the theory
and leads to huge enhancement of the corresponding oblique
corrections for any value of $m_{SUSY}$. To begin with, let us neglect
$\tilde{t}_L \tilde{t}_R$ mixing. In this case we have with good
accuracy:  $m^2_{\tilde{t}_L} - m^2_{\tilde{b}_L} = m_t^2$ (in what
follows we will often designate $\tilde{b}_L$ as $\tilde{b}$). For
$(\tilde{t}_L, \tilde{b}_L)$ loop insertion into $W$-boson propagator
we get the result, proportional to $(m^2_{\tilde{t}_L} -
m^2_{\tilde{b}_L})^2 / m_W^2 m^2_{SUSY} \approx 16(m_W/m_{SUSY})^2$
since $m_t \approx 2m_W$. That is why the $(\tilde{t}, \tilde{b})$ 
oblique
corrections are responsible for the main part of SUSY corrections
to gluon-free observables. (Let us note that the existence of
the terms $\sim (m_t)^4$ in the SUSY radiative corrections was
observed long ago \cite{7}. Now when we know that top quark is very
heavy and from the direct searches it follows that most of SUSY
partners should be much heavier than $M_Z /2$ we can claim that the
stop-sbottom contribution to the oblique corrections dominates over
the other SUSY contributions to gluon-free observables.) The term
$\sim m_t^4/m^2_W m^2_{SUSY}$ comes from $\Sigma_W(0)$, while
the terms proportional to $\Sigma'(0)$ produce the corrections of
the order of $m_t^2 /m^2_{SUSY}$. Higher order derivatives of self
energies are suppressed as $(m_{W,Z} /m_{SUSY})^2$ and should be
omitted.  Calculating the vector boson self energies and keeping the
first two terms in the $k^2$ expansion, we get the following
expressions for the SUSY contributions to the quantities $V_i$, which
are directly related to the physical observables ($V_m$ -- to $m_W$,
$V_A$ -- to $g_A$ and $V_R$ -- to the ratio $g_V/g_A$, see \cite{6}):
\begin{equation}
\delta_{SUSY} V_A = \frac{1}{m_Z^2} g(m_{\tilde{t}_L}, m_{\tilde{b}})
\;\; ,
\label{1}
\end{equation}

\begin{equation}
\delta_{SUSY} V_R = \delta_{SUSY} V_A + \frac{1}{3} Y_L
\ln(\frac{m^2_{\tilde{t}_L}}{m^2_{\tilde{b}}}) \;\; ,
\label{2}
\end{equation}

\begin{equation}
\delta_{SUSY} V_m = \delta_{SUSY} V_A + \frac{2}{3} Y_L s^2
\ln(\frac{m^2_{\tilde{t}_L}}{m^2_{\tilde{b}}})
+(\frac{c^2 -s^2}{3}) h(m_{\tilde{t}_L}, m_{\tilde{b}}) \;\; ,
\label{3}
\end{equation}
where $Y_L$ is the left doublet hypercharge, $Y_L = Q_t + Q_b$, $s
\equiv \sin\theta$, $c\equiv \cos\theta$, $\sin^2\theta \cos^2\theta
= \pi\bar{\alpha}(m_Z)/(\sqrt{2} G_{\mu}m_Z^2)$ ($\theta$ is the
electroweak mixing angle) and the functions $g$ and $h$ are standard
functions which describe the scalar particle contributions to the
vector boson self energies at one loop:
\begin{equation}
g(m_1, m_2) = m_1^2 +m_2^2 -2\frac{m_1^2 m_2^2}{m_1^2 - m_2^2} \ln
(\frac{m_1^2}{m_2^2}) \;\; ,
\label{5}
\end{equation}
\begin{equation}
h(m_1, m_2) = -\frac{5}{3} + 
\frac{4m_1^2 m_2^2}{(m_1^2 -m_2^2)^2} + \nonumber \\
 \frac{(m_1^2 +m_2^2)(m_1^4 -4m_1^2 m_2^2 +m_2^4)}{(m_1^2 -m_2^2)^3}
\ln (\frac{m_1^2}{m_2^2}) \;\; .
\label{6}
\end{equation}

For SUSY partners with  masses larger than $m_t$, 
$m_{\tilde{b}} \gg m_t$,
the leading term in (\ref{1}) -- (\ref{3}) is universal and
originates from $\Sigma_W(0)$ and is given by the function $g$:
\begin{equation}
\delta_{SUSY} V_i = \frac{1}{3} \frac{(m^2_{\tilde{t_L}} -
m^2_{\tilde{b}})^2}{m^2_Z m^2_{\tilde{b}}} \;\; .
\label{7}
\end{equation}

For light SUSY partners $m_{\tilde{b}} \sim m_t$ we should not expand
$\delta_{SUSY} V_i$ over $(m_{\tilde{t_L}}^2 -
m^2_{\tilde{b}})/m^2_{\tilde{b}}$ but use expressions (\ref{1}) --
(\ref{3}) to describe the electroweak corrections. In Figures 1 - 3
the dependences of the functions $\delta_{SUSY} V_i$ on
$m_{\tilde{b}}$ are shown. For top quark mass we use $m_t=175GeV$.
The horizontal lines show the difference
of the experimental values of $V_i$ and their Standard Model fit. The
solid lines correspond to the central values while the dotted ones
correspond to the one standard deviation corridor. From the Figures 
we see that
for $m_{\tilde{b}} \geq 200$ GeV SUSY corrections are not essential
and the inclusion of SUSY partners will not spoil the Standard Model
fit of the data.
\vspace{0.5cm}

\setlength{\unitlength}{0.240900pt}
\ifx\plotpoint\undefined\newsavebox{\plotpoint}\fi
\begin{picture}(1500,900)(0,0)
\font\gnuplot=cmr10 at 10pt
\sbox{\plotpoint}{\rule[-0.200pt]{0.400pt}{0.400pt}}%
\put(176.0,304.0){\rule[-0.200pt]{303.534pt}{0.400pt}}
\put(176.0,113.0){\rule[-0.200pt]{4.818pt}{0.400pt}}
\put(154,113){\makebox(0,0)[r]{-1}}
\put(1416.0,113.0){\rule[-0.200pt]{4.818pt}{0.400pt}}
\put(176.0,209.0){\rule[-0.200pt]{4.818pt}{0.400pt}}
\put(154,209){\makebox(0,0)[r]{-0.5}}
\put(1416.0,209.0){\rule[-0.200pt]{4.818pt}{0.400pt}}
\put(176.0,304.0){\rule[-0.200pt]{4.818pt}{0.400pt}}
\put(154,304){\makebox(0,0)[r]{0}}
\put(1416.0,304.0){\rule[-0.200pt]{4.818pt}{0.400pt}}
\put(176.0,400.0){\rule[-0.200pt]{4.818pt}{0.400pt}}
\put(154,400){\makebox(0,0)[r]{0.5}}
\put(1416.0,400.0){\rule[-0.200pt]{4.818pt}{0.400pt}}
\put(176.0,495.0){\rule[-0.200pt]{4.818pt}{0.400pt}}
\put(154,495){\makebox(0,0)[r]{1}}
\put(1416.0,495.0){\rule[-0.200pt]{4.818pt}{0.400pt}}
\put(176.0,591.0){\rule[-0.200pt]{4.818pt}{0.400pt}}
\put(154,591){\makebox(0,0)[r]{1.5}}
\put(1416.0,591.0){\rule[-0.200pt]{4.818pt}{0.400pt}}
\put(176.0,686.0){\rule[-0.200pt]{4.818pt}{0.400pt}}
\put(154,686){\makebox(0,0)[r]{2}}
\put(1416.0,686.0){\rule[-0.200pt]{4.818pt}{0.400pt}}
\put(176.0,782.0){\rule[-0.200pt]{4.818pt}{0.400pt}}
\put(154,782){\makebox(0,0)[r]{2.5}}
\put(1416.0,782.0){\rule[-0.200pt]{4.818pt}{0.400pt}}
\put(176.0,877.0){\rule[-0.200pt]{4.818pt}{0.400pt}}
\put(154,877){\makebox(0,0)[r]{3}}
\put(1416.0,877.0){\rule[-0.200pt]{4.818pt}{0.400pt}}
\put(175.0,113.0){\rule[-0.200pt]{0.400pt}{4.818pt}}
\put(175,68){\makebox(0,0){0}}
\put(175.0,857.0){\rule[-0.200pt]{0.400pt}{4.818pt}}
\put(301.0,113.0){\rule[-0.200pt]{0.400pt}{4.818pt}}
\put(301,68){\makebox(0,0){100}}
\put(301.0,857.0){\rule[-0.200pt]{0.400pt}{4.818pt}}
\put(427.0,113.0){\rule[-0.200pt]{0.400pt}{4.818pt}}
\put(427,68){\makebox(0,0){200}}
\put(427.0,857.0){\rule[-0.200pt]{0.400pt}{4.818pt}}
\put(553.0,113.0){\rule[-0.200pt]{0.400pt}{4.818pt}}
\put(553,68){\makebox(0,0){300}}
\put(553.0,857.0){\rule[-0.200pt]{0.400pt}{4.818pt}}
\put(679.0,113.0){\rule[-0.200pt]{0.400pt}{4.818pt}}
\put(679,68){\makebox(0,0){400}}
\put(679.0,857.0){\rule[-0.200pt]{0.400pt}{4.818pt}}
\put(805.0,113.0){\rule[-0.200pt]{0.400pt}{4.818pt}}
\put(805,68){\makebox(0,0){500}}
\put(805.0,857.0){\rule[-0.200pt]{0.400pt}{4.818pt}}
\put(931.0,113.0){\rule[-0.200pt]{0.400pt}{4.818pt}}
\put(931,68){\makebox(0,0){600}}
\put(931.0,857.0){\rule[-0.200pt]{0.400pt}{4.818pt}}
\put(1058.0,113.0){\rule[-0.200pt]{0.400pt}{4.818pt}}
\put(1058,68){\makebox(0,0){700}}
\put(1058.0,857.0){\rule[-0.200pt]{0.400pt}{4.818pt}}
\put(1184.0,113.0){\rule[-0.200pt]{0.400pt}{4.818pt}}
\put(1184,68){\makebox(0,0){800}}
\put(1184.0,857.0){\rule[-0.200pt]{0.400pt}{4.818pt}}
\put(1310.0,113.0){\rule[-0.200pt]{0.400pt}{4.818pt}}
\put(1310,68){\makebox(0,0){900}}
\put(1310.0,857.0){\rule[-0.200pt]{0.400pt}{4.818pt}}
\put(1436.0,113.0){\rule[-0.200pt]{0.400pt}{4.818pt}}
\put(1436,68){\makebox(0,0){1000}}
\put(1436.0,857.0){\rule[-0.200pt]{0.400pt}{4.818pt}}
\put(176.0,113.0){\rule[-0.200pt]{303.534pt}{0.400pt}}
\put(1436.0,113.0){\rule[-0.200pt]{0.400pt}{184.048pt}}
\put(176.0,877.0){\rule[-0.200pt]{303.534pt}{0.400pt}}
\put(806,23){\makebox(0,0){$m_{\tilde{b}}$}}
\put(176.0,113.0){\rule[-0.200pt]{0.400pt}{184.048pt}}
\put(176,231){\usebox{\plotpoint}}
\put(176.0,231.0){\rule[-0.200pt]{303.534pt}{0.400pt}}
\put(176,319){\usebox{\plotpoint}}
\put(176.00,319.00){\usebox{\plotpoint}}
\put(196.76,319.00){\usebox{\plotpoint}}
\multiput(201,319)(20.756,0.000){0}{\usebox{\plotpoint}}
\put(217.51,319.00){\usebox{\plotpoint}}
\put(238.27,319.00){\usebox{\plotpoint}}
\multiput(240,319)(20.756,0.000){0}{\usebox{\plotpoint}}
\put(259.02,319.00){\usebox{\plotpoint}}
\multiput(265,319)(20.756,0.000){0}{\usebox{\plotpoint}}
\put(279.78,319.00){\usebox{\plotpoint}}
\put(300.53,319.00){\usebox{\plotpoint}}
\multiput(303,319)(20.756,0.000){0}{\usebox{\plotpoint}}
\put(321.29,319.00){\usebox{\plotpoint}}
\multiput(329,319)(20.756,0.000){0}{\usebox{\plotpoint}}
\put(342.04,319.00){\usebox{\plotpoint}}
\put(362.80,319.00){\usebox{\plotpoint}}
\multiput(367,319)(20.756,0.000){0}{\usebox{\plotpoint}}
\put(383.55,319.00){\usebox{\plotpoint}}
\put(404.31,319.00){\usebox{\plotpoint}}
\multiput(405,319)(20.756,0.000){0}{\usebox{\plotpoint}}
\put(425.07,319.00){\usebox{\plotpoint}}
\multiput(431,319)(20.756,0.000){0}{\usebox{\plotpoint}}
\put(445.82,319.00){\usebox{\plotpoint}}
\put(466.58,319.00){\usebox{\plotpoint}}
\multiput(469,319)(20.756,0.000){0}{\usebox{\plotpoint}}
\put(487.33,319.00){\usebox{\plotpoint}}
\multiput(494,319)(20.756,0.000){0}{\usebox{\plotpoint}}
\put(508.09,319.00){\usebox{\plotpoint}}
\put(528.84,319.00){\usebox{\plotpoint}}
\multiput(532,319)(20.756,0.000){0}{\usebox{\plotpoint}}
\put(549.60,319.00){\usebox{\plotpoint}}
\put(570.35,319.00){\usebox{\plotpoint}}
\multiput(571,319)(20.756,0.000){0}{\usebox{\plotpoint}}
\put(591.11,319.00){\usebox{\plotpoint}}
\multiput(596,319)(20.756,0.000){0}{\usebox{\plotpoint}}
\put(611.87,319.00){\usebox{\plotpoint}}
\put(632.62,319.00){\usebox{\plotpoint}}
\multiput(634,319)(20.756,0.000){0}{\usebox{\plotpoint}}
\put(653.38,319.00){\usebox{\plotpoint}}
\multiput(660,319)(20.756,0.000){0}{\usebox{\plotpoint}}
\put(674.13,319.00){\usebox{\plotpoint}}
\put(694.89,319.00){\usebox{\plotpoint}}
\multiput(698,319)(20.756,0.000){0}{\usebox{\plotpoint}}
\put(715.64,319.00){\usebox{\plotpoint}}
\multiput(723,319)(20.756,0.000){0}{\usebox{\plotpoint}}
\put(736.40,319.00){\usebox{\plotpoint}}
\put(757.15,319.00){\usebox{\plotpoint}}
\multiput(761,319)(20.756,0.000){0}{\usebox{\plotpoint}}
\put(777.91,319.00){\usebox{\plotpoint}}
\put(798.66,319.00){\usebox{\plotpoint}}
\multiput(800,319)(20.756,0.000){0}{\usebox{\plotpoint}}
\put(819.42,319.00){\usebox{\plotpoint}}
\multiput(825,319)(20.756,0.000){0}{\usebox{\plotpoint}}
\put(840.18,319.00){\usebox{\plotpoint}}
\put(860.93,319.00){\usebox{\plotpoint}}
\multiput(863,319)(20.756,0.000){0}{\usebox{\plotpoint}}
\put(881.69,319.00){\usebox{\plotpoint}}
\multiput(889,319)(20.756,0.000){0}{\usebox{\plotpoint}}
\put(902.44,319.00){\usebox{\plotpoint}}
\put(923.20,319.00){\usebox{\plotpoint}}
\multiput(927,319)(20.756,0.000){0}{\usebox{\plotpoint}}
\put(943.95,319.00){\usebox{\plotpoint}}
\put(964.71,319.00){\usebox{\plotpoint}}
\multiput(965,319)(20.756,0.000){0}{\usebox{\plotpoint}}
\put(985.46,319.00){\usebox{\plotpoint}}
\multiput(991,319)(20.756,0.000){0}{\usebox{\plotpoint}}
\put(1006.22,319.00){\usebox{\plotpoint}}
\put(1026.98,319.00){\usebox{\plotpoint}}
\multiput(1029,319)(20.756,0.000){0}{\usebox{\plotpoint}}
\put(1047.73,319.00){\usebox{\plotpoint}}
\multiput(1054,319)(20.756,0.000){0}{\usebox{\plotpoint}}
\put(1068.49,319.00){\usebox{\plotpoint}}
\put(1089.24,319.00){\usebox{\plotpoint}}
\multiput(1092,319)(20.756,0.000){0}{\usebox{\plotpoint}}
\put(1110.00,319.00){\usebox{\plotpoint}}
\put(1130.75,319.00){\usebox{\plotpoint}}
\multiput(1131,319)(20.756,0.000){0}{\usebox{\plotpoint}}
\put(1151.51,319.00){\usebox{\plotpoint}}
\multiput(1156,319)(20.756,0.000){0}{\usebox{\plotpoint}}
\put(1172.26,319.00){\usebox{\plotpoint}}
\put(1193.02,319.00){\usebox{\plotpoint}}
\multiput(1194,319)(20.756,0.000){0}{\usebox{\plotpoint}}
\put(1213.77,319.00){\usebox{\plotpoint}}
\multiput(1220,319)(20.756,0.000){0}{\usebox{\plotpoint}}
\put(1234.53,319.00){\usebox{\plotpoint}}
\put(1255.29,319.00){\usebox{\plotpoint}}
\multiput(1258,319)(20.756,0.000){0}{\usebox{\plotpoint}}
\put(1276.04,319.00){\usebox{\plotpoint}}
\multiput(1283,319)(20.756,0.000){0}{\usebox{\plotpoint}}
\put(1296.80,319.00){\usebox{\plotpoint}}
\put(1317.55,319.00){\usebox{\plotpoint}}
\multiput(1321,319)(20.756,0.000){0}{\usebox{\plotpoint}}
\put(1338.31,319.00){\usebox{\plotpoint}}
\put(1359.06,319.00){\usebox{\plotpoint}}
\multiput(1360,319)(20.756,0.000){0}{\usebox{\plotpoint}}
\put(1379.82,319.00){\usebox{\plotpoint}}
\multiput(1385,319)(20.756,0.000){0}{\usebox{\plotpoint}}
\put(1400.57,319.00){\usebox{\plotpoint}}
\put(1421.33,319.00){\usebox{\plotpoint}}
\multiput(1423,319)(20.756,0.000){0}{\usebox{\plotpoint}}
\put(1436,319){\usebox{\plotpoint}}
\put(176,144){\usebox{\plotpoint}}
\put(176.00,144.00){\usebox{\plotpoint}}
\put(196.76,144.00){\usebox{\plotpoint}}
\multiput(201,144)(20.756,0.000){0}{\usebox{\plotpoint}}
\put(217.51,144.00){\usebox{\plotpoint}}
\put(238.27,144.00){\usebox{\plotpoint}}
\multiput(240,144)(20.756,0.000){0}{\usebox{\plotpoint}}
\put(259.02,144.00){\usebox{\plotpoint}}
\multiput(265,144)(20.756,0.000){0}{\usebox{\plotpoint}}
\put(279.78,144.00){\usebox{\plotpoint}}
\put(300.53,144.00){\usebox{\plotpoint}}
\multiput(303,144)(20.756,0.000){0}{\usebox{\plotpoint}}
\put(321.29,144.00){\usebox{\plotpoint}}
\multiput(329,144)(20.756,0.000){0}{\usebox{\plotpoint}}
\put(342.04,144.00){\usebox{\plotpoint}}
\put(362.80,144.00){\usebox{\plotpoint}}
\multiput(367,144)(20.756,0.000){0}{\usebox{\plotpoint}}
\put(383.55,144.00){\usebox{\plotpoint}}
\put(404.31,144.00){\usebox{\plotpoint}}
\multiput(405,144)(20.756,0.000){0}{\usebox{\plotpoint}}
\put(425.07,144.00){\usebox{\plotpoint}}
\multiput(431,144)(20.756,0.000){0}{\usebox{\plotpoint}}
\put(445.82,144.00){\usebox{\plotpoint}}
\put(466.58,144.00){\usebox{\plotpoint}}
\multiput(469,144)(20.756,0.000){0}{\usebox{\plotpoint}}
\put(487.33,144.00){\usebox{\plotpoint}}
\multiput(494,144)(20.756,0.000){0}{\usebox{\plotpoint}}
\put(508.09,144.00){\usebox{\plotpoint}}
\put(528.84,144.00){\usebox{\plotpoint}}
\multiput(532,144)(20.756,0.000){0}{\usebox{\plotpoint}}
\put(549.60,144.00){\usebox{\plotpoint}}
\put(570.35,144.00){\usebox{\plotpoint}}
\multiput(571,144)(20.756,0.000){0}{\usebox{\plotpoint}}
\put(591.11,144.00){\usebox{\plotpoint}}
\multiput(596,144)(20.756,0.000){0}{\usebox{\plotpoint}}
\put(611.87,144.00){\usebox{\plotpoint}}
\put(632.62,144.00){\usebox{\plotpoint}}
\multiput(634,144)(20.756,0.000){0}{\usebox{\plotpoint}}
\put(653.38,144.00){\usebox{\plotpoint}}
\multiput(660,144)(20.756,0.000){0}{\usebox{\plotpoint}}
\put(674.13,144.00){\usebox{\plotpoint}}
\put(694.89,144.00){\usebox{\plotpoint}}
\multiput(698,144)(20.756,0.000){0}{\usebox{\plotpoint}}
\put(715.64,144.00){\usebox{\plotpoint}}
\multiput(723,144)(20.756,0.000){0}{\usebox{\plotpoint}}
\put(736.40,144.00){\usebox{\plotpoint}}
\put(757.15,144.00){\usebox{\plotpoint}}
\multiput(761,144)(20.756,0.000){0}{\usebox{\plotpoint}}
\put(777.91,144.00){\usebox{\plotpoint}}
\put(798.66,144.00){\usebox{\plotpoint}}
\multiput(800,144)(20.756,0.000){0}{\usebox{\plotpoint}}
\put(819.42,144.00){\usebox{\plotpoint}}
\multiput(825,144)(20.756,0.000){0}{\usebox{\plotpoint}}
\put(840.18,144.00){\usebox{\plotpoint}}
\put(860.93,144.00){\usebox{\plotpoint}}
\multiput(863,144)(20.756,0.000){0}{\usebox{\plotpoint}}
\put(881.69,144.00){\usebox{\plotpoint}}
\multiput(889,144)(20.756,0.000){0}{\usebox{\plotpoint}}
\put(902.44,144.00){\usebox{\plotpoint}}
\put(923.20,144.00){\usebox{\plotpoint}}
\multiput(927,144)(20.756,0.000){0}{\usebox{\plotpoint}}
\put(943.95,144.00){\usebox{\plotpoint}}
\put(964.71,144.00){\usebox{\plotpoint}}
\multiput(965,144)(20.756,0.000){0}{\usebox{\plotpoint}}
\put(985.46,144.00){\usebox{\plotpoint}}
\multiput(991,144)(20.756,0.000){0}{\usebox{\plotpoint}}
\put(1006.22,144.00){\usebox{\plotpoint}}
\put(1026.98,144.00){\usebox{\plotpoint}}
\multiput(1029,144)(20.756,0.000){0}{\usebox{\plotpoint}}
\put(1047.73,144.00){\usebox{\plotpoint}}
\multiput(1054,144)(20.756,0.000){0}{\usebox{\plotpoint}}
\put(1068.49,144.00){\usebox{\plotpoint}}
\put(1089.24,144.00){\usebox{\plotpoint}}
\multiput(1092,144)(20.756,0.000){0}{\usebox{\plotpoint}}
\put(1110.00,144.00){\usebox{\plotpoint}}
\put(1130.75,144.00){\usebox{\plotpoint}}
\multiput(1131,144)(20.756,0.000){0}{\usebox{\plotpoint}}
\put(1151.51,144.00){\usebox{\plotpoint}}
\multiput(1156,144)(20.756,0.000){0}{\usebox{\plotpoint}}
\put(1172.26,144.00){\usebox{\plotpoint}}
\put(1193.02,144.00){\usebox{\plotpoint}}
\multiput(1194,144)(20.756,0.000){0}{\usebox{\plotpoint}}
\put(1213.77,144.00){\usebox{\plotpoint}}
\multiput(1220,144)(20.756,0.000){0}{\usebox{\plotpoint}}
\put(1234.53,144.00){\usebox{\plotpoint}}
\put(1255.29,144.00){\usebox{\plotpoint}}
\multiput(1258,144)(20.756,0.000){0}{\usebox{\plotpoint}}
\put(1276.04,144.00){\usebox{\plotpoint}}
\multiput(1283,144)(20.756,0.000){0}{\usebox{\plotpoint}}
\put(1296.80,144.00){\usebox{\plotpoint}}
\put(1317.55,144.00){\usebox{\plotpoint}}
\multiput(1321,144)(20.756,0.000){0}{\usebox{\plotpoint}}
\put(1338.31,144.00){\usebox{\plotpoint}}
\put(1359.06,144.00){\usebox{\plotpoint}}
\multiput(1360,144)(20.756,0.000){0}{\usebox{\plotpoint}}
\put(1379.82,144.00){\usebox{\plotpoint}}
\multiput(1385,144)(20.756,0.000){0}{\usebox{\plotpoint}}
\put(1400.57,144.00){\usebox{\plotpoint}}
\put(1421.33,144.00){\usebox{\plotpoint}}
\multiput(1423,144)(20.756,0.000){0}{\usebox{\plotpoint}}
\put(1436,144){\usebox{\plotpoint}}
\put(1306,812){\makebox(0,0)[r]{$V_A$}}
\put(1328.0,812.0){\rule[-0.200pt]{15.899pt}{0.400pt}}
\put(238,806){\usebox{\plotpoint}}
\multiput(238.58,799.91)(0.497,-1.723){59}{\rule{0.120pt}{1.468pt}}
\multiput(237.17,802.95)(31.000,-102.954){2}{\rule{0.400pt}{0.734pt}}
\multiput(269.58,695.23)(0.497,-1.320){61}{\rule{0.120pt}{1.150pt}}
\multiput(268.17,697.61)(32.000,-81.613){2}{\rule{0.400pt}{0.575pt}}
\multiput(301.58,612.16)(0.497,-1.037){59}{\rule{0.120pt}{0.926pt}}
\multiput(300.17,614.08)(31.000,-62.078){2}{\rule{0.400pt}{0.463pt}}
\multiput(332.58,549.04)(0.497,-0.767){61}{\rule{0.120pt}{0.713pt}}
\multiput(331.17,550.52)(32.000,-47.521){2}{\rule{0.400pt}{0.356pt}}
\multiput(364.58,500.60)(0.497,-0.597){59}{\rule{0.120pt}{0.577pt}}
\multiput(363.17,501.80)(31.000,-35.802){2}{\rule{0.400pt}{0.289pt}}
\multiput(395.00,464.92)(0.551,-0.497){55}{\rule{0.541pt}{0.120pt}}
\multiput(395.00,465.17)(30.876,-29.000){2}{\rule{0.271pt}{0.400pt}}
\multiput(427.00,435.92)(0.729,-0.496){41}{\rule{0.682pt}{0.120pt}}
\multiput(427.00,436.17)(30.585,-22.000){2}{\rule{0.341pt}{0.400pt}}
\multiput(459.00,413.92)(0.919,-0.495){31}{\rule{0.829pt}{0.119pt}}
\multiput(459.00,414.17)(29.279,-17.000){2}{\rule{0.415pt}{0.400pt}}
\multiput(490.00,396.92)(1.158,-0.494){25}{\rule{1.014pt}{0.119pt}}
\multiput(490.00,397.17)(29.895,-14.000){2}{\rule{0.507pt}{0.400pt}}
\multiput(522.00,382.92)(1.439,-0.492){19}{\rule{1.227pt}{0.118pt}}
\multiput(522.00,383.17)(28.453,-11.000){2}{\rule{0.614pt}{0.400pt}}
\multiput(553.00,371.93)(1.834,-0.489){15}{\rule{1.522pt}{0.118pt}}
\multiput(553.00,372.17)(28.841,-9.000){2}{\rule{0.761pt}{0.400pt}}
\multiput(585.00,362.93)(2.323,-0.485){11}{\rule{1.871pt}{0.117pt}}
\multiput(585.00,363.17)(27.116,-7.000){2}{\rule{0.936pt}{0.400pt}}
\multiput(616.00,355.93)(2.841,-0.482){9}{\rule{2.233pt}{0.116pt}}
\multiput(616.00,356.17)(27.365,-6.000){2}{\rule{1.117pt}{0.400pt}}
\multiput(648.00,349.93)(3.382,-0.477){7}{\rule{2.580pt}{0.115pt}}
\multiput(648.00,350.17)(25.645,-5.000){2}{\rule{1.290pt}{0.400pt}}
\multiput(679.00,344.93)(3.493,-0.477){7}{\rule{2.660pt}{0.115pt}}
\multiput(679.00,345.17)(26.479,-5.000){2}{\rule{1.330pt}{0.400pt}}
\multiput(711.00,339.94)(4.429,-0.468){5}{\rule{3.200pt}{0.113pt}}
\multiput(711.00,340.17)(24.358,-4.000){2}{\rule{1.600pt}{0.400pt}}
\multiput(742.00,335.95)(6.937,-0.447){3}{\rule{4.367pt}{0.108pt}}
\multiput(742.00,336.17)(22.937,-3.000){2}{\rule{2.183pt}{0.400pt}}
\multiput(774.00,332.95)(6.714,-0.447){3}{\rule{4.233pt}{0.108pt}}
\multiput(774.00,333.17)(22.214,-3.000){2}{\rule{2.117pt}{0.400pt}}
\put(805,329.17){\rule{6.500pt}{0.400pt}}
\multiput(805.00,330.17)(18.509,-2.000){2}{\rule{3.250pt}{0.400pt}}
\put(837,327.17){\rule{6.300pt}{0.400pt}}
\multiput(837.00,328.17)(17.924,-2.000){2}{\rule{3.150pt}{0.400pt}}
\put(868,325.17){\rule{6.500pt}{0.400pt}}
\multiput(868.00,326.17)(18.509,-2.000){2}{\rule{3.250pt}{0.400pt}}
\put(900,323.17){\rule{6.300pt}{0.400pt}}
\multiput(900.00,324.17)(17.924,-2.000){2}{\rule{3.150pt}{0.400pt}}
\put(931,321.67){\rule{7.709pt}{0.400pt}}
\multiput(931.00,322.17)(16.000,-1.000){2}{\rule{3.854pt}{0.400pt}}
\put(963,320.67){\rule{7.709pt}{0.400pt}}
\multiput(963.00,321.17)(16.000,-1.000){2}{\rule{3.854pt}{0.400pt}}
\put(995,319.17){\rule{6.300pt}{0.400pt}}
\multiput(995.00,320.17)(17.924,-2.000){2}{\rule{3.150pt}{0.400pt}}
\put(1026,317.67){\rule{7.709pt}{0.400pt}}
\multiput(1026.00,318.17)(16.000,-1.000){2}{\rule{3.854pt}{0.400pt}}
\put(1058,316.67){\rule{7.468pt}{0.400pt}}
\multiput(1058.00,317.17)(15.500,-1.000){2}{\rule{3.734pt}{0.400pt}}
\put(1121,315.67){\rule{7.468pt}{0.400pt}}
\multiput(1121.00,316.17)(15.500,-1.000){2}{\rule{3.734pt}{0.400pt}}
\put(1152,314.67){\rule{7.709pt}{0.400pt}}
\multiput(1152.00,315.17)(16.000,-1.000){2}{\rule{3.854pt}{0.400pt}}
\put(1184,313.67){\rule{7.468pt}{0.400pt}}
\multiput(1184.00,314.17)(15.500,-1.000){2}{\rule{3.734pt}{0.400pt}}
\put(1089.0,317.0){\rule[-0.200pt]{7.709pt}{0.400pt}}
\put(1247,312.67){\rule{7.468pt}{0.400pt}}
\multiput(1247.00,313.17)(15.500,-1.000){2}{\rule{3.734pt}{0.400pt}}
\put(1215.0,314.0){\rule[-0.200pt]{7.709pt}{0.400pt}}
\put(1310,311.67){\rule{7.468pt}{0.400pt}}
\multiput(1310.00,312.17)(15.500,-1.000){2}{\rule{3.734pt}{0.400pt}}
\put(1278.0,313.0){\rule[-0.200pt]{7.709pt}{0.400pt}}
\put(1341.0,312.0){\rule[-0.200pt]{15.177pt}{0.400pt}}
\end{picture}

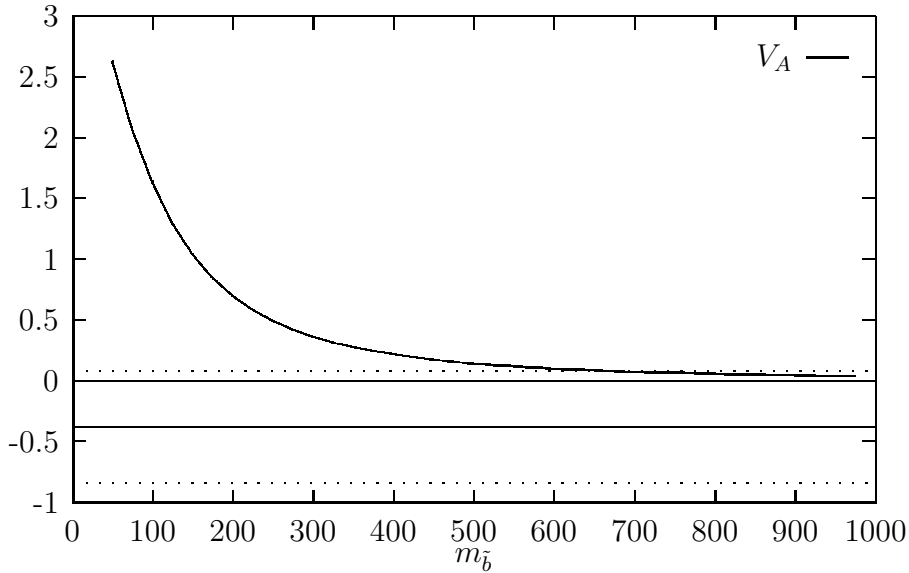
Figure 1: The dependence of the function $\delta_{SUSY}V_A$ on $m_{\tilde b}$.
      The horizontal lines  show the difference between experimental
      value of $V_A$ and its fit in the Standard Model,  $V_A^{exp}-V_A^{theor}=-0.38(46)$. The solid line corresponds to the central value,
while dotted -- to one standard deviation corridor.


\setlength{\unitlength}{0.240900pt}
\ifx\plotpoint\undefined\newsavebox{\plotpoint}\fi
\begin{picture}(1500,900)(0,0)
\font\gnuplot=cmr10 at 10pt
\gnuplot
\sbox{\plotpoint}{\rule[-0.200pt]{0.400pt}{0.400pt}}%
\put(176.0,222.0){\rule[-0.200pt]{303.534pt}{0.400pt}}
\put(176.0,113.0){\rule[-0.200pt]{4.818pt}{0.400pt}}
\put(154,113){\makebox(0,0)[r]{-0.5}}
\put(1416.0,113.0){\rule[-0.200pt]{4.818pt}{0.400pt}}
\put(176.0,222.0){\rule[-0.200pt]{4.818pt}{0.400pt}}
\put(154,222){\makebox(0,0)[r]{0}}
\put(1416.0,222.0){\rule[-0.200pt]{4.818pt}{0.400pt}}
\put(176.0,331.0){\rule[-0.200pt]{4.818pt}{0.400pt}}
\put(154,331){\makebox(0,0)[r]{0.5}}
\put(1416.0,331.0){\rule[-0.200pt]{4.818pt}{0.400pt}}
\put(176.0,440.0){\rule[-0.200pt]{4.818pt}{0.400pt}}
\put(154,440){\makebox(0,0)[r]{1}}
\put(1416.0,440.0){\rule[-0.200pt]{4.818pt}{0.400pt}}
\put(176.0,550.0){\rule[-0.200pt]{4.818pt}{0.400pt}}
\put(154,550){\makebox(0,0)[r]{1.5}}
\put(1416.0,550.0){\rule[-0.200pt]{4.818pt}{0.400pt}}
\put(176.0,659.0){\rule[-0.200pt]{4.818pt}{0.400pt}}
\put(154,659){\makebox(0,0)[r]{2}}
\put(1416.0,659.0){\rule[-0.200pt]{4.818pt}{0.400pt}}
\put(176.0,768.0){\rule[-0.200pt]{4.818pt}{0.400pt}}
\put(154,768){\makebox(0,0)[r]{2.5}}
\put(1416.0,768.0){\rule[-0.200pt]{4.818pt}{0.400pt}}
\put(176.0,877.0){\rule[-0.200pt]{4.818pt}{0.400pt}}
\put(154,877){\makebox(0,0)[r]{3}}
\put(1416.0,877.0){\rule[-0.200pt]{4.818pt}{0.400pt}}
\put(175.0,113.0){\rule[-0.200pt]{0.400pt}{4.818pt}}
\put(175,68){\makebox(0,0){0}}
\put(175.0,857.0){\rule[-0.200pt]{0.400pt}{4.818pt}}
\put(301.0,113.0){\rule[-0.200pt]{0.400pt}{4.818pt}}
\put(301,68){\makebox(0,0){100}}
\put(301.0,857.0){\rule[-0.200pt]{0.400pt}{4.818pt}}
\put(427.0,113.0){\rule[-0.200pt]{0.400pt}{4.818pt}}
\put(427,68){\makebox(0,0){200}}
\put(427.0,857.0){\rule[-0.200pt]{0.400pt}{4.818pt}}
\put(553.0,113.0){\rule[-0.200pt]{0.400pt}{4.818pt}}
\put(553,68){\makebox(0,0){300}}
\put(553.0,857.0){\rule[-0.200pt]{0.400pt}{4.818pt}}
\put(679.0,113.0){\rule[-0.200pt]{0.400pt}{4.818pt}}
\put(679,68){\makebox(0,0){400}}
\put(679.0,857.0){\rule[-0.200pt]{0.400pt}{4.818pt}}
\put(805.0,113.0){\rule[-0.200pt]{0.400pt}{4.818pt}}
\put(805,68){\makebox(0,0){500}}
\put(805.0,857.0){\rule[-0.200pt]{0.400pt}{4.818pt}}
\put(931.0,113.0){\rule[-0.200pt]{0.400pt}{4.818pt}}
\put(931,68){\makebox(0,0){600}}
\put(931.0,857.0){\rule[-0.200pt]{0.400pt}{4.818pt}}
\put(1058.0,113.0){\rule[-0.200pt]{0.400pt}{4.818pt}}
\put(1058,68){\makebox(0,0){700}}
\put(1058.0,857.0){\rule[-0.200pt]{0.400pt}{4.818pt}}
\put(1184.0,113.0){\rule[-0.200pt]{0.400pt}{4.818pt}}
\put(1184,68){\makebox(0,0){800}}
\put(1184.0,857.0){\rule[-0.200pt]{0.400pt}{4.818pt}}
\put(1310.0,113.0){\rule[-0.200pt]{0.400pt}{4.818pt}}
\put(1310,68){\makebox(0,0){900}}
\put(1310.0,857.0){\rule[-0.200pt]{0.400pt}{4.818pt}}
\put(1436.0,113.0){\rule[-0.200pt]{0.400pt}{4.818pt}}
\put(1436,68){\makebox(0,0){1000}}
\put(1436.0,857.0){\rule[-0.200pt]{0.400pt}{4.818pt}}
\put(176.0,113.0){\rule[-0.200pt]{303.534pt}{0.400pt}}
\put(1436.0,113.0){\rule[-0.200pt]{0.400pt}{184.048pt}}
\put(176.0,877.0){\rule[-0.200pt]{303.534pt}{0.400pt}}
\put(806,23){\makebox(0,0){$m_{\tilde{b}}$}}
\put(176.0,113.0){\rule[-0.200pt]{0.400pt}{184.048pt}}
\put(176,222){\usebox{\plotpoint}}
\put(176.0,222.0){\rule[-0.200pt]{303.534pt}{0.400pt}}
\put(176,309){\usebox{\plotpoint}}
\put(176.00,309.00){\usebox{\plotpoint}}
\put(196.76,309.00){\usebox{\plotpoint}}
\multiput(201,309)(20.756,0.000){0}{\usebox{\plotpoint}}
\put(217.51,309.00){\usebox{\plotpoint}}
\put(238.27,309.00){\usebox{\plotpoint}}
\multiput(240,309)(20.756,0.000){0}{\usebox{\plotpoint}}
\put(259.02,309.00){\usebox{\plotpoint}}
\multiput(265,309)(20.756,0.000){0}{\usebox{\plotpoint}}
\put(279.78,309.00){\usebox{\plotpoint}}
\put(300.53,309.00){\usebox{\plotpoint}}
\multiput(303,309)(20.756,0.000){0}{\usebox{\plotpoint}}
\put(321.29,309.00){\usebox{\plotpoint}}
\multiput(329,309)(20.756,0.000){0}{\usebox{\plotpoint}}
\put(342.04,309.00){\usebox{\plotpoint}}
\put(362.80,309.00){\usebox{\plotpoint}}
\multiput(367,309)(20.756,0.000){0}{\usebox{\plotpoint}}
\put(383.55,309.00){\usebox{\plotpoint}}
\put(404.31,309.00){\usebox{\plotpoint}}
\multiput(405,309)(20.756,0.000){0}{\usebox{\plotpoint}}
\put(425.07,309.00){\usebox{\plotpoint}}
\multiput(431,309)(20.756,0.000){0}{\usebox{\plotpoint}}
\put(445.82,309.00){\usebox{\plotpoint}}
\put(466.58,309.00){\usebox{\plotpoint}}
\multiput(469,309)(20.756,0.000){0}{\usebox{\plotpoint}}
\put(487.33,309.00){\usebox{\plotpoint}}
\multiput(494,309)(20.756,0.000){0}{\usebox{\plotpoint}}
\put(508.09,309.00){\usebox{\plotpoint}}
\put(528.84,309.00){\usebox{\plotpoint}}
\multiput(532,309)(20.756,0.000){0}{\usebox{\plotpoint}}
\put(549.60,309.00){\usebox{\plotpoint}}
\put(570.35,309.00){\usebox{\plotpoint}}
\multiput(571,309)(20.756,0.000){0}{\usebox{\plotpoint}}
\put(591.11,309.00){\usebox{\plotpoint}}
\multiput(596,309)(20.756,0.000){0}{\usebox{\plotpoint}}
\put(611.87,309.00){\usebox{\plotpoint}}
\put(632.62,309.00){\usebox{\plotpoint}}
\multiput(634,309)(20.756,0.000){0}{\usebox{\plotpoint}}
\put(653.38,309.00){\usebox{\plotpoint}}
\multiput(660,309)(20.756,0.000){0}{\usebox{\plotpoint}}
\put(674.13,309.00){\usebox{\plotpoint}}
\put(694.89,309.00){\usebox{\plotpoint}}
\multiput(698,309)(20.756,0.000){0}{\usebox{\plotpoint}}
\put(715.64,309.00){\usebox{\plotpoint}}
\multiput(723,309)(20.756,0.000){0}{\usebox{\plotpoint}}
\put(736.40,309.00){\usebox{\plotpoint}}
\put(757.15,309.00){\usebox{\plotpoint}}
\multiput(761,309)(20.756,0.000){0}{\usebox{\plotpoint}}
\put(777.91,309.00){\usebox{\plotpoint}}
\put(798.66,309.00){\usebox{\plotpoint}}
\multiput(800,309)(20.756,0.000){0}{\usebox{\plotpoint}}
\put(819.42,309.00){\usebox{\plotpoint}}
\multiput(825,309)(20.756,0.000){0}{\usebox{\plotpoint}}
\put(840.18,309.00){\usebox{\plotpoint}}
\put(860.93,309.00){\usebox{\plotpoint}}
\multiput(863,309)(20.756,0.000){0}{\usebox{\plotpoint}}
\put(881.69,309.00){\usebox{\plotpoint}}
\multiput(889,309)(20.756,0.000){0}{\usebox{\plotpoint}}
\put(902.44,309.00){\usebox{\plotpoint}}
\put(923.20,309.00){\usebox{\plotpoint}}
\multiput(927,309)(20.756,0.000){0}{\usebox{\plotpoint}}
\put(943.95,309.00){\usebox{\plotpoint}}
\put(964.71,309.00){\usebox{\plotpoint}}
\multiput(965,309)(20.756,0.000){0}{\usebox{\plotpoint}}
\put(985.46,309.00){\usebox{\plotpoint}}
\multiput(991,309)(20.756,0.000){0}{\usebox{\plotpoint}}
\put(1006.22,309.00){\usebox{\plotpoint}}
\put(1026.98,309.00){\usebox{\plotpoint}}
\multiput(1029,309)(20.756,0.000){0}{\usebox{\plotpoint}}
\put(1047.73,309.00){\usebox{\plotpoint}}
\multiput(1054,309)(20.756,0.000){0}{\usebox{\plotpoint}}
\put(1068.49,309.00){\usebox{\plotpoint}}
\put(1089.24,309.00){\usebox{\plotpoint}}
\multiput(1092,309)(20.756,0.000){0}{\usebox{\plotpoint}}
\put(1110.00,309.00){\usebox{\plotpoint}}
\put(1130.75,309.00){\usebox{\plotpoint}}
\multiput(1131,309)(20.756,0.000){0}{\usebox{\plotpoint}}
\put(1151.51,309.00){\usebox{\plotpoint}}
\multiput(1156,309)(20.756,0.000){0}{\usebox{\plotpoint}}
\put(1172.26,309.00){\usebox{\plotpoint}}
\put(1193.02,309.00){\usebox{\plotpoint}}
\multiput(1194,309)(20.756,0.000){0}{\usebox{\plotpoint}}
\put(1213.77,309.00){\usebox{\plotpoint}}
\multiput(1220,309)(20.756,0.000){0}{\usebox{\plotpoint}}
\put(1234.53,309.00){\usebox{\plotpoint}}
\put(1255.29,309.00){\usebox{\plotpoint}}
\multiput(1258,309)(20.756,0.000){0}{\usebox{\plotpoint}}
\put(1276.04,309.00){\usebox{\plotpoint}}
\multiput(1283,309)(20.756,0.000){0}{\usebox{\plotpoint}}
\put(1296.80,309.00){\usebox{\plotpoint}}
\put(1317.55,309.00){\usebox{\plotpoint}}
\multiput(1321,309)(20.756,0.000){0}{\usebox{\plotpoint}}
\put(1338.31,309.00){\usebox{\plotpoint}}
\put(1359.06,309.00){\usebox{\plotpoint}}
\multiput(1360,309)(20.756,0.000){0}{\usebox{\plotpoint}}
\put(1379.82,309.00){\usebox{\plotpoint}}
\multiput(1385,309)(20.756,0.000){0}{\usebox{\plotpoint}}
\put(1400.57,309.00){\usebox{\plotpoint}}
\put(1421.33,309.00){\usebox{\plotpoint}}
\multiput(1423,309)(20.756,0.000){0}{\usebox{\plotpoint}}
\put(1436,309){\usebox{\plotpoint}}
\put(176,135){\usebox{\plotpoint}}
\put(176.00,135.00){\usebox{\plotpoint}}
\put(196.76,135.00){\usebox{\plotpoint}}
\multiput(201,135)(20.756,0.000){0}{\usebox{\plotpoint}}
\put(217.51,135.00){\usebox{\plotpoint}}
\put(238.27,135.00){\usebox{\plotpoint}}
\multiput(240,135)(20.756,0.000){0}{\usebox{\plotpoint}}
\put(259.02,135.00){\usebox{\plotpoint}}
\multiput(265,135)(20.756,0.000){0}{\usebox{\plotpoint}}
\put(279.78,135.00){\usebox{\plotpoint}}
\put(300.53,135.00){\usebox{\plotpoint}}
\multiput(303,135)(20.756,0.000){0}{\usebox{\plotpoint}}
\put(321.29,135.00){\usebox{\plotpoint}}
\multiput(329,135)(20.756,0.000){0}{\usebox{\plotpoint}}
\put(342.04,135.00){\usebox{\plotpoint}}
\put(362.80,135.00){\usebox{\plotpoint}}
\multiput(367,135)(20.756,0.000){0}{\usebox{\plotpoint}}
\put(383.55,135.00){\usebox{\plotpoint}}
\put(404.31,135.00){\usebox{\plotpoint}}
\multiput(405,135)(20.756,0.000){0}{\usebox{\plotpoint}}
\put(425.07,135.00){\usebox{\plotpoint}}
\multiput(431,135)(20.756,0.000){0}{\usebox{\plotpoint}}
\put(445.82,135.00){\usebox{\plotpoint}}
\put(466.58,135.00){\usebox{\plotpoint}}
\multiput(469,135)(20.756,0.000){0}{\usebox{\plotpoint}}
\put(487.33,135.00){\usebox{\plotpoint}}
\multiput(494,135)(20.756,0.000){0}{\usebox{\plotpoint}}
\put(508.09,135.00){\usebox{\plotpoint}}
\put(528.84,135.00){\usebox{\plotpoint}}
\multiput(532,135)(20.756,0.000){0}{\usebox{\plotpoint}}
\put(549.60,135.00){\usebox{\plotpoint}}
\put(570.35,135.00){\usebox{\plotpoint}}
\multiput(571,135)(20.756,0.000){0}{\usebox{\plotpoint}}
\put(591.11,135.00){\usebox{\plotpoint}}
\multiput(596,135)(20.756,0.000){0}{\usebox{\plotpoint}}
\put(611.87,135.00){\usebox{\plotpoint}}
\put(632.62,135.00){\usebox{\plotpoint}}
\multiput(634,135)(20.756,0.000){0}{\usebox{\plotpoint}}
\put(653.38,135.00){\usebox{\plotpoint}}
\multiput(660,135)(20.756,0.000){0}{\usebox{\plotpoint}}
\put(674.13,135.00){\usebox{\plotpoint}}
\put(694.89,135.00){\usebox{\plotpoint}}
\multiput(698,135)(20.756,0.000){0}{\usebox{\plotpoint}}
\put(715.64,135.00){\usebox{\plotpoint}}
\multiput(723,135)(20.756,0.000){0}{\usebox{\plotpoint}}
\put(736.40,135.00){\usebox{\plotpoint}}
\put(757.15,135.00){\usebox{\plotpoint}}
\multiput(761,135)(20.756,0.000){0}{\usebox{\plotpoint}}
\put(777.91,135.00){\usebox{\plotpoint}}
\put(798.66,135.00){\usebox{\plotpoint}}
\multiput(800,135)(20.756,0.000){0}{\usebox{\plotpoint}}
\put(819.42,135.00){\usebox{\plotpoint}}
\multiput(825,135)(20.756,0.000){0}{\usebox{\plotpoint}}
\put(840.18,135.00){\usebox{\plotpoint}}
\put(860.93,135.00){\usebox{\plotpoint}}
\multiput(863,135)(20.756,0.000){0}{\usebox{\plotpoint}}
\put(881.69,135.00){\usebox{\plotpoint}}
\multiput(889,135)(20.756,0.000){0}{\usebox{\plotpoint}}
\put(902.44,135.00){\usebox{\plotpoint}}
\put(923.20,135.00){\usebox{\plotpoint}}
\multiput(927,135)(20.756,0.000){0}{\usebox{\plotpoint}}
\put(943.95,135.00){\usebox{\plotpoint}}
\put(964.71,135.00){\usebox{\plotpoint}}
\multiput(965,135)(20.756,0.000){0}{\usebox{\plotpoint}}
\put(985.46,135.00){\usebox{\plotpoint}}
\multiput(991,135)(20.756,0.000){0}{\usebox{\plotpoint}}
\put(1006.22,135.00){\usebox{\plotpoint}}
\put(1026.98,135.00){\usebox{\plotpoint}}
\multiput(1029,135)(20.756,0.000){0}{\usebox{\plotpoint}}
\put(1047.73,135.00){\usebox{\plotpoint}}
\multiput(1054,135)(20.756,0.000){0}{\usebox{\plotpoint}}
\put(1068.49,135.00){\usebox{\plotpoint}}
\put(1089.24,135.00){\usebox{\plotpoint}}
\multiput(1092,135)(20.756,0.000){0}{\usebox{\plotpoint}}
\put(1110.00,135.00){\usebox{\plotpoint}}
\put(1130.75,135.00){\usebox{\plotpoint}}
\multiput(1131,135)(20.756,0.000){0}{\usebox{\plotpoint}}
\put(1151.51,135.00){\usebox{\plotpoint}}
\multiput(1156,135)(20.756,0.000){0}{\usebox{\plotpoint}}
\put(1172.26,135.00){\usebox{\plotpoint}}
\put(1193.02,135.00){\usebox{\plotpoint}}
\multiput(1194,135)(20.756,0.000){0}{\usebox{\plotpoint}}
\put(1213.77,135.00){\usebox{\plotpoint}}
\multiput(1220,135)(20.756,0.000){0}{\usebox{\plotpoint}}
\put(1234.53,135.00){\usebox{\plotpoint}}
\put(1255.29,135.00){\usebox{\plotpoint}}
\multiput(1258,135)(20.756,0.000){0}{\usebox{\plotpoint}}
\put(1276.04,135.00){\usebox{\plotpoint}}
\multiput(1283,135)(20.756,0.000){0}{\usebox{\plotpoint}}
\put(1296.80,135.00){\usebox{\plotpoint}}
\put(1317.55,135.00){\usebox{\plotpoint}}
\multiput(1321,135)(20.756,0.000){0}{\usebox{\plotpoint}}
\put(1338.31,135.00){\usebox{\plotpoint}}
\put(1359.06,135.00){\usebox{\plotpoint}}
\multiput(1360,135)(20.756,0.000){0}{\usebox{\plotpoint}}
\put(1379.82,135.00){\usebox{\plotpoint}}
\multiput(1385,135)(20.756,0.000){0}{\usebox{\plotpoint}}
\put(1400.57,135.00){\usebox{\plotpoint}}
\put(1421.33,135.00){\usebox{\plotpoint}}
\multiput(1423,135)(20.756,0.000){0}{\usebox{\plotpoint}}
\put(1436,135){\usebox{\plotpoint}}
\put(1306,812){\makebox(0,0)[r]{$V_R$}}
\put(1328.0,812.0){\rule[-0.200pt]{15.899pt}{0.400pt}}
\put(238,858){\usebox{\plotpoint}}
\multiput(238.58,850.19)(0.497,-2.245){59}{\rule{0.120pt}{1.881pt}}
\multiput(237.17,854.10)(31.000,-134.097){2}{\rule{0.400pt}{0.940pt}}
\multiput(269.58,714.03)(0.497,-1.684){61}{\rule{0.120pt}{1.438pt}}
\multiput(268.17,717.02)(32.000,-104.016){2}{\rule{0.400pt}{0.719pt}}
\multiput(301.58,608.19)(0.497,-1.331){59}{\rule{0.120pt}{1.158pt}}
\multiput(300.17,610.60)(31.000,-79.596){2}{\rule{0.400pt}{0.579pt}}
\multiput(332.58,527.42)(0.497,-0.957){61}{\rule{0.120pt}{0.863pt}}
\multiput(331.17,529.21)(32.000,-59.210){2}{\rule{0.400pt}{0.431pt}}
\multiput(364.58,467.12)(0.497,-0.743){59}{\rule{0.120pt}{0.694pt}}
\multiput(363.17,468.56)(31.000,-44.561){2}{\rule{0.400pt}{0.347pt}}
\multiput(395.58,421.77)(0.497,-0.546){61}{\rule{0.120pt}{0.538pt}}
\multiput(394.17,422.88)(32.000,-33.884){2}{\rule{0.400pt}{0.269pt}}
\multiput(427.00,387.92)(0.571,-0.497){53}{\rule{0.557pt}{0.120pt}}
\multiput(427.00,388.17)(30.844,-28.000){2}{\rule{0.279pt}{0.400pt}}
\multiput(459.00,359.92)(0.740,-0.496){39}{\rule{0.690pt}{0.119pt}}
\multiput(459.00,360.17)(29.567,-21.000){2}{\rule{0.345pt}{0.400pt}}
\multiput(490.00,338.92)(0.895,-0.495){33}{\rule{0.811pt}{0.119pt}}
\multiput(490.00,339.17)(30.316,-18.000){2}{\rule{0.406pt}{0.400pt}}
\multiput(522.00,320.92)(1.210,-0.493){23}{\rule{1.054pt}{0.119pt}}
\multiput(522.00,321.17)(28.813,-13.000){2}{\rule{0.527pt}{0.400pt}}
\multiput(553.00,307.92)(1.358,-0.492){21}{\rule{1.167pt}{0.119pt}}
\multiput(553.00,308.17)(29.579,-12.000){2}{\rule{0.583pt}{0.400pt}}
\multiput(585.00,295.93)(1.776,-0.489){15}{\rule{1.478pt}{0.118pt}}
\multiput(585.00,296.17)(27.933,-9.000){2}{\rule{0.739pt}{0.400pt}}
\multiput(616.00,286.93)(2.079,-0.488){13}{\rule{1.700pt}{0.117pt}}
\multiput(616.00,287.17)(28.472,-8.000){2}{\rule{0.850pt}{0.400pt}}
\multiput(648.00,278.93)(2.751,-0.482){9}{\rule{2.167pt}{0.116pt}}
\multiput(648.00,279.17)(26.503,-6.000){2}{\rule{1.083pt}{0.400pt}}
\multiput(679.00,272.93)(2.841,-0.482){9}{\rule{2.233pt}{0.116pt}}
\multiput(679.00,273.17)(27.365,-6.000){2}{\rule{1.117pt}{0.400pt}}
\multiput(711.00,266.94)(4.429,-0.468){5}{\rule{3.200pt}{0.113pt}}
\multiput(711.00,267.17)(24.358,-4.000){2}{\rule{1.600pt}{0.400pt}}
\multiput(742.00,262.94)(4.575,-0.468){5}{\rule{3.300pt}{0.113pt}}
\multiput(742.00,263.17)(25.151,-4.000){2}{\rule{1.650pt}{0.400pt}}
\multiput(774.00,258.94)(4.429,-0.468){5}{\rule{3.200pt}{0.113pt}}
\multiput(774.00,259.17)(24.358,-4.000){2}{\rule{1.600pt}{0.400pt}}
\multiput(805.00,254.95)(6.937,-0.447){3}{\rule{4.367pt}{0.108pt}}
\multiput(805.00,255.17)(22.937,-3.000){2}{\rule{2.183pt}{0.400pt}}
\put(837,251.17){\rule{6.300pt}{0.400pt}}
\multiput(837.00,252.17)(17.924,-2.000){2}{\rule{3.150pt}{0.400pt}}
\multiput(868.00,249.95)(6.937,-0.447){3}{\rule{4.367pt}{0.108pt}}
\multiput(868.00,250.17)(22.937,-3.000){2}{\rule{2.183pt}{0.400pt}}
\put(900,246.17){\rule{6.300pt}{0.400pt}}
\multiput(900.00,247.17)(17.924,-2.000){2}{\rule{3.150pt}{0.400pt}}
\put(931,244.17){\rule{6.500pt}{0.400pt}}
\multiput(931.00,245.17)(18.509,-2.000){2}{\rule{3.250pt}{0.400pt}}
\put(963,242.67){\rule{7.709pt}{0.400pt}}
\multiput(963.00,243.17)(16.000,-1.000){2}{\rule{3.854pt}{0.400pt}}
\put(995,241.17){\rule{6.300pt}{0.400pt}}
\multiput(995.00,242.17)(17.924,-2.000){2}{\rule{3.150pt}{0.400pt}}
\put(1026,239.67){\rule{7.709pt}{0.400pt}}
\multiput(1026.00,240.17)(16.000,-1.000){2}{\rule{3.854pt}{0.400pt}}
\put(1058,238.67){\rule{7.468pt}{0.400pt}}
\multiput(1058.00,239.17)(15.500,-1.000){2}{\rule{3.734pt}{0.400pt}}
\put(1089,237.67){\rule{7.709pt}{0.400pt}}
\multiput(1089.00,238.17)(16.000,-1.000){2}{\rule{3.854pt}{0.400pt}}
\put(1121,236.67){\rule{7.468pt}{0.400pt}}
\multiput(1121.00,237.17)(15.500,-1.000){2}{\rule{3.734pt}{0.400pt}}
\put(1152,235.67){\rule{7.709pt}{0.400pt}}
\multiput(1152.00,236.17)(16.000,-1.000){2}{\rule{3.854pt}{0.400pt}}
\put(1184,234.67){\rule{7.468pt}{0.400pt}}
\multiput(1184.00,235.17)(15.500,-1.000){2}{\rule{3.734pt}{0.400pt}}
\put(1215,233.67){\rule{7.709pt}{0.400pt}}
\multiput(1215.00,234.17)(16.000,-1.000){2}{\rule{3.854pt}{0.400pt}}
\put(1278,232.67){\rule{7.709pt}{0.400pt}}
\multiput(1278.00,233.17)(16.000,-1.000){2}{\rule{3.854pt}{0.400pt}}
\put(1247.0,234.0){\rule[-0.200pt]{7.468pt}{0.400pt}}
\put(1341,231.67){\rule{7.709pt}{0.400pt}}
\multiput(1341.00,232.17)(16.000,-1.000){2}{\rule{3.854pt}{0.400pt}}
\put(1310.0,233.0){\rule[-0.200pt]{7.468pt}{0.400pt}}
\put(1404,230.67){\rule{7.709pt}{0.400pt}}
\multiput(1404.00,231.17)(16.000,-1.000){2}{\rule{3.854pt}{0.400pt}}
\put(1373.0,232.0){\rule[-0.200pt]{7.468pt}{0.400pt}}
\end{picture}

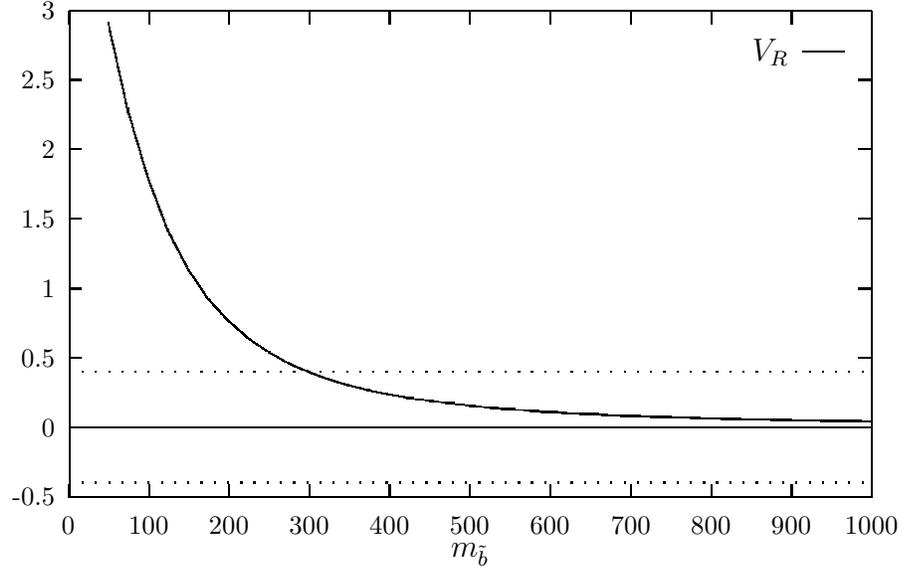
Figure 2: The same as Figure 1 for $V_R$. $V_R^{exp}-V_R^{theor}=-0.0(4)$ was used.
\vspace{8mm}

\setlength{\unitlength}{0.240900pt}
\ifx\plotpoint\undefined\newsavebox{\plotpoint}\fi
\begin{picture}(1500,900)(0,0)
\font\gnuplot=cmr10 at 10pt
\gnuplot
\sbox{\plotpoint}{\rule[-0.200pt]{0.400pt}{0.400pt}}%
\put(176.0,222.0){\rule[-0.200pt]{303.534pt}{0.400pt}}
\put(176.0,113.0){\rule[-0.200pt]{4.818pt}{0.400pt}}
\put(154,113){\makebox(0,0)[r]{-0.5}}
\put(1416.0,113.0){\rule[-0.200pt]{4.818pt}{0.400pt}}
\put(176.0,222.0){\rule[-0.200pt]{4.818pt}{0.400pt}}
\put(154,222){\makebox(0,0)[r]{0}}
\put(1416.0,222.0){\rule[-0.200pt]{4.818pt}{0.400pt}}
\put(176.0,331.0){\rule[-0.200pt]{4.818pt}{0.400pt}}
\put(154,331){\makebox(0,0)[r]{0.5}}
\put(1416.0,331.0){\rule[-0.200pt]{4.818pt}{0.400pt}}
\put(176.0,440.0){\rule[-0.200pt]{4.818pt}{0.400pt}}
\put(154,440){\makebox(0,0)[r]{1}}
\put(1416.0,440.0){\rule[-0.200pt]{4.818pt}{0.400pt}}
\put(176.0,550.0){\rule[-0.200pt]{4.818pt}{0.400pt}}
\put(154,550){\makebox(0,0)[r]{1.5}}
\put(1416.0,550.0){\rule[-0.200pt]{4.818pt}{0.400pt}}
\put(176.0,659.0){\rule[-0.200pt]{4.818pt}{0.400pt}}
\put(154,659){\makebox(0,0)[r]{2}}
\put(1416.0,659.0){\rule[-0.200pt]{4.818pt}{0.400pt}}
\put(176.0,768.0){\rule[-0.200pt]{4.818pt}{0.400pt}}
\put(154,768){\makebox(0,0)[r]{2.5}}
\put(1416.0,768.0){\rule[-0.200pt]{4.818pt}{0.400pt}}
\put(176.0,877.0){\rule[-0.200pt]{4.818pt}{0.400pt}}
\put(154,877){\makebox(0,0)[r]{3}}
\put(1416.0,877.0){\rule[-0.200pt]{4.818pt}{0.400pt}}
\put(175.0,113.0){\rule[-0.200pt]{0.400pt}{4.818pt}}
\put(175,68){\makebox(0,0){0}}
\put(175.0,857.0){\rule[-0.200pt]{0.400pt}{4.818pt}}
\put(301.0,113.0){\rule[-0.200pt]{0.400pt}{4.818pt}}
\put(301,68){\makebox(0,0){100}}
\put(301.0,857.0){\rule[-0.200pt]{0.400pt}{4.818pt}}
\put(427.0,113.0){\rule[-0.200pt]{0.400pt}{4.818pt}}
\put(427,68){\makebox(0,0){200}}
\put(427.0,857.0){\rule[-0.200pt]{0.400pt}{4.818pt}}
\put(553.0,113.0){\rule[-0.200pt]{0.400pt}{4.818pt}}
\put(553,68){\makebox(0,0){300}}
\put(553.0,857.0){\rule[-0.200pt]{0.400pt}{4.818pt}}
\put(679.0,113.0){\rule[-0.200pt]{0.400pt}{4.818pt}}
\put(679,68){\makebox(0,0){400}}
\put(679.0,857.0){\rule[-0.200pt]{0.400pt}{4.818pt}}
\put(805.0,113.0){\rule[-0.200pt]{0.400pt}{4.818pt}}
\put(805,68){\makebox(0,0){500}}
\put(805.0,857.0){\rule[-0.200pt]{0.400pt}{4.818pt}}
\put(931.0,113.0){\rule[-0.200pt]{0.400pt}{4.818pt}}
\put(931,68){\makebox(0,0){600}}
\put(931.0,857.0){\rule[-0.200pt]{0.400pt}{4.818pt}}
\put(1058.0,113.0){\rule[-0.200pt]{0.400pt}{4.818pt}}
\put(1058,68){\makebox(0,0){700}}
\put(1058.0,857.0){\rule[-0.200pt]{0.400pt}{4.818pt}}
\put(1184.0,113.0){\rule[-0.200pt]{0.400pt}{4.818pt}}
\put(1184,68){\makebox(0,0){800}}
\put(1184.0,857.0){\rule[-0.200pt]{0.400pt}{4.818pt}}
\put(1310.0,113.0){\rule[-0.200pt]{0.400pt}{4.818pt}}
\put(1310,68){\makebox(0,0){900}}
\put(1310.0,857.0){\rule[-0.200pt]{0.400pt}{4.818pt}}
\put(1436.0,113.0){\rule[-0.200pt]{0.400pt}{4.818pt}}
\put(1436,68){\makebox(0,0){1000}}
\put(1436.0,857.0){\rule[-0.200pt]{0.400pt}{4.818pt}}
\put(176.0,113.0){\rule[-0.200pt]{303.534pt}{0.400pt}}
\put(1436.0,113.0){\rule[-0.200pt]{0.400pt}{184.048pt}}
\put(176.0,877.0){\rule[-0.200pt]{303.534pt}{0.400pt}}
\put(806,23){\makebox(0,0){$m_{\tilde{b}}$}}
\put(176.0,113.0){\rule[-0.200pt]{0.400pt}{184.048pt}}
\put(176,320){\usebox{\plotpoint}}
\put(176.0,320.0){\rule[-0.200pt]{303.534pt}{0.400pt}}
\put(176,462){\usebox{\plotpoint}}
\put(176.00,462.00){\usebox{\plotpoint}}
\put(196.76,462.00){\usebox{\plotpoint}}
\multiput(201,462)(20.756,0.000){0}{\usebox{\plotpoint}}
\put(217.51,462.00){\usebox{\plotpoint}}
\put(238.27,462.00){\usebox{\plotpoint}}
\multiput(240,462)(20.756,0.000){0}{\usebox{\plotpoint}}
\put(259.02,462.00){\usebox{\plotpoint}}
\multiput(265,462)(20.756,0.000){0}{\usebox{\plotpoint}}
\put(279.78,462.00){\usebox{\plotpoint}}
\put(300.53,462.00){\usebox{\plotpoint}}
\multiput(303,462)(20.756,0.000){0}{\usebox{\plotpoint}}
\put(321.29,462.00){\usebox{\plotpoint}}
\multiput(329,462)(20.756,0.000){0}{\usebox{\plotpoint}}
\put(342.04,462.00){\usebox{\plotpoint}}
\put(362.80,462.00){\usebox{\plotpoint}}
\multiput(367,462)(20.756,0.000){0}{\usebox{\plotpoint}}
\put(383.55,462.00){\usebox{\plotpoint}}
\put(404.31,462.00){\usebox{\plotpoint}}
\multiput(405,462)(20.756,0.000){0}{\usebox{\plotpoint}}
\put(425.07,462.00){\usebox{\plotpoint}}
\multiput(431,462)(20.756,0.000){0}{\usebox{\plotpoint}}
\put(445.82,462.00){\usebox{\plotpoint}}
\put(466.58,462.00){\usebox{\plotpoint}}
\multiput(469,462)(20.756,0.000){0}{\usebox{\plotpoint}}
\put(487.33,462.00){\usebox{\plotpoint}}
\multiput(494,462)(20.756,0.000){0}{\usebox{\plotpoint}}
\put(508.09,462.00){\usebox{\plotpoint}}
\put(528.84,462.00){\usebox{\plotpoint}}
\multiput(532,462)(20.756,0.000){0}{\usebox{\plotpoint}}
\put(549.60,462.00){\usebox{\plotpoint}}
\put(570.35,462.00){\usebox{\plotpoint}}
\multiput(571,462)(20.756,0.000){0}{\usebox{\plotpoint}}
\put(591.11,462.00){\usebox{\plotpoint}}
\multiput(596,462)(20.756,0.000){0}{\usebox{\plotpoint}}
\put(611.87,462.00){\usebox{\plotpoint}}
\put(632.62,462.00){\usebox{\plotpoint}}
\multiput(634,462)(20.756,0.000){0}{\usebox{\plotpoint}}
\put(653.38,462.00){\usebox{\plotpoint}}
\multiput(660,462)(20.756,0.000){0}{\usebox{\plotpoint}}
\put(674.13,462.00){\usebox{\plotpoint}}
\put(694.89,462.00){\usebox{\plotpoint}}
\multiput(698,462)(20.756,0.000){0}{\usebox{\plotpoint}}
\put(715.64,462.00){\usebox{\plotpoint}}
\multiput(723,462)(20.756,0.000){0}{\usebox{\plotpoint}}
\put(736.40,462.00){\usebox{\plotpoint}}
\put(757.15,462.00){\usebox{\plotpoint}}
\multiput(761,462)(20.756,0.000){0}{\usebox{\plotpoint}}
\put(777.91,462.00){\usebox{\plotpoint}}
\put(798.66,462.00){\usebox{\plotpoint}}
\multiput(800,462)(20.756,0.000){0}{\usebox{\plotpoint}}
\put(819.42,462.00){\usebox{\plotpoint}}
\multiput(825,462)(20.756,0.000){0}{\usebox{\plotpoint}}
\put(840.18,462.00){\usebox{\plotpoint}}
\put(860.93,462.00){\usebox{\plotpoint}}
\multiput(863,462)(20.756,0.000){0}{\usebox{\plotpoint}}
\put(881.69,462.00){\usebox{\plotpoint}}
\multiput(889,462)(20.756,0.000){0}{\usebox{\plotpoint}}
\put(902.44,462.00){\usebox{\plotpoint}}
\put(923.20,462.00){\usebox{\plotpoint}}
\multiput(927,462)(20.756,0.000){0}{\usebox{\plotpoint}}
\put(943.95,462.00){\usebox{\plotpoint}}
\put(964.71,462.00){\usebox{\plotpoint}}
\multiput(965,462)(20.756,0.000){0}{\usebox{\plotpoint}}
\put(985.46,462.00){\usebox{\plotpoint}}
\multiput(991,462)(20.756,0.000){0}{\usebox{\plotpoint}}
\put(1006.22,462.00){\usebox{\plotpoint}}
\put(1026.98,462.00){\usebox{\plotpoint}}
\multiput(1029,462)(20.756,0.000){0}{\usebox{\plotpoint}}
\put(1047.73,462.00){\usebox{\plotpoint}}
\multiput(1054,462)(20.756,0.000){0}{\usebox{\plotpoint}}
\put(1068.49,462.00){\usebox{\plotpoint}}
\put(1089.24,462.00){\usebox{\plotpoint}}
\multiput(1092,462)(20.756,0.000){0}{\usebox{\plotpoint}}
\put(1110.00,462.00){\usebox{\plotpoint}}
\put(1130.75,462.00){\usebox{\plotpoint}}
\multiput(1131,462)(20.756,0.000){0}{\usebox{\plotpoint}}
\put(1151.51,462.00){\usebox{\plotpoint}}
\multiput(1156,462)(20.756,0.000){0}{\usebox{\plotpoint}}
\put(1172.26,462.00){\usebox{\plotpoint}}
\put(1193.02,462.00){\usebox{\plotpoint}}
\multiput(1194,462)(20.756,0.000){0}{\usebox{\plotpoint}}
\put(1213.77,462.00){\usebox{\plotpoint}}
\multiput(1220,462)(20.756,0.000){0}{\usebox{\plotpoint}}
\put(1234.53,462.00){\usebox{\plotpoint}}
\put(1255.29,462.00){\usebox{\plotpoint}}
\multiput(1258,462)(20.756,0.000){0}{\usebox{\plotpoint}}
\put(1276.04,462.00){\usebox{\plotpoint}}
\multiput(1283,462)(20.756,0.000){0}{\usebox{\plotpoint}}
\put(1296.80,462.00){\usebox{\plotpoint}}
\put(1317.55,462.00){\usebox{\plotpoint}}
\multiput(1321,462)(20.756,0.000){0}{\usebox{\plotpoint}}
\put(1338.31,462.00){\usebox{\plotpoint}}
\put(1359.06,462.00){\usebox{\plotpoint}}
\multiput(1360,462)(20.756,0.000){0}{\usebox{\plotpoint}}
\put(1379.82,462.00){\usebox{\plotpoint}}
\multiput(1385,462)(20.756,0.000){0}{\usebox{\plotpoint}}
\put(1400.57,462.00){\usebox{\plotpoint}}
\put(1421.33,462.00){\usebox{\plotpoint}}
\multiput(1423,462)(20.756,0.000){0}{\usebox{\plotpoint}}
\put(1436,462){\usebox{\plotpoint}}
\put(176,178){\usebox{\plotpoint}}
\put(176.00,178.00){\usebox{\plotpoint}}
\put(196.76,178.00){\usebox{\plotpoint}}
\multiput(201,178)(20.756,0.000){0}{\usebox{\plotpoint}}
\put(217.51,178.00){\usebox{\plotpoint}}
\put(238.27,178.00){\usebox{\plotpoint}}
\multiput(240,178)(20.756,0.000){0}{\usebox{\plotpoint}}
\put(259.02,178.00){\usebox{\plotpoint}}
\multiput(265,178)(20.756,0.000){0}{\usebox{\plotpoint}}
\put(279.78,178.00){\usebox{\plotpoint}}
\put(300.53,178.00){\usebox{\plotpoint}}
\multiput(303,178)(20.756,0.000){0}{\usebox{\plotpoint}}
\put(321.29,178.00){\usebox{\plotpoint}}
\multiput(329,178)(20.756,0.000){0}{\usebox{\plotpoint}}
\put(342.04,178.00){\usebox{\plotpoint}}
\put(362.80,178.00){\usebox{\plotpoint}}
\multiput(367,178)(20.756,0.000){0}{\usebox{\plotpoint}}
\put(383.55,178.00){\usebox{\plotpoint}}
\put(404.31,178.00){\usebox{\plotpoint}}
\multiput(405,178)(20.756,0.000){0}{\usebox{\plotpoint}}
\put(425.07,178.00){\usebox{\plotpoint}}
\multiput(431,178)(20.756,0.000){0}{\usebox{\plotpoint}}
\put(445.82,178.00){\usebox{\plotpoint}}
\put(466.58,178.00){\usebox{\plotpoint}}
\multiput(469,178)(20.756,0.000){0}{\usebox{\plotpoint}}
\put(487.33,178.00){\usebox{\plotpoint}}
\multiput(494,178)(20.756,0.000){0}{\usebox{\plotpoint}}
\put(508.09,178.00){\usebox{\plotpoint}}
\put(528.84,178.00){\usebox{\plotpoint}}
\multiput(532,178)(20.756,0.000){0}{\usebox{\plotpoint}}
\put(549.60,178.00){\usebox{\plotpoint}}
\put(570.35,178.00){\usebox{\plotpoint}}
\multiput(571,178)(20.756,0.000){0}{\usebox{\plotpoint}}
\put(591.11,178.00){\usebox{\plotpoint}}
\multiput(596,178)(20.756,0.000){0}{\usebox{\plotpoint}}
\put(611.87,178.00){\usebox{\plotpoint}}
\put(632.62,178.00){\usebox{\plotpoint}}
\multiput(634,178)(20.756,0.000){0}{\usebox{\plotpoint}}
\put(653.38,178.00){\usebox{\plotpoint}}
\multiput(660,178)(20.756,0.000){0}{\usebox{\plotpoint}}
\put(674.13,178.00){\usebox{\plotpoint}}
\put(694.89,178.00){\usebox{\plotpoint}}
\multiput(698,178)(20.756,0.000){0}{\usebox{\plotpoint}}
\put(715.64,178.00){\usebox{\plotpoint}}
\multiput(723,178)(20.756,0.000){0}{\usebox{\plotpoint}}
\put(736.40,178.00){\usebox{\plotpoint}}
\put(757.15,178.00){\usebox{\plotpoint}}
\multiput(761,178)(20.756,0.000){0}{\usebox{\plotpoint}}
\put(777.91,178.00){\usebox{\plotpoint}}
\put(798.66,178.00){\usebox{\plotpoint}}
\multiput(800,178)(20.756,0.000){0}{\usebox{\plotpoint}}
\put(819.42,178.00){\usebox{\plotpoint}}
\multiput(825,178)(20.756,0.000){0}{\usebox{\plotpoint}}
\put(840.18,178.00){\usebox{\plotpoint}}
\put(860.93,178.00){\usebox{\plotpoint}}
\multiput(863,178)(20.756,0.000){0}{\usebox{\plotpoint}}
\put(881.69,178.00){\usebox{\plotpoint}}
\multiput(889,178)(20.756,0.000){0}{\usebox{\plotpoint}}
\put(902.44,178.00){\usebox{\plotpoint}}
\put(923.20,178.00){\usebox{\plotpoint}}
\multiput(927,178)(20.756,0.000){0}{\usebox{\plotpoint}}
\put(943.95,178.00){\usebox{\plotpoint}}
\put(964.71,178.00){\usebox{\plotpoint}}
\multiput(965,178)(20.756,0.000){0}{\usebox{\plotpoint}}
\put(985.46,178.00){\usebox{\plotpoint}}
\multiput(991,178)(20.756,0.000){0}{\usebox{\plotpoint}}
\put(1006.22,178.00){\usebox{\plotpoint}}
\put(1026.98,178.00){\usebox{\plotpoint}}
\multiput(1029,178)(20.756,0.000){0}{\usebox{\plotpoint}}
\put(1047.73,178.00){\usebox{\plotpoint}}
\multiput(1054,178)(20.756,0.000){0}{\usebox{\plotpoint}}
\put(1068.49,178.00){\usebox{\plotpoint}}
\put(1089.24,178.00){\usebox{\plotpoint}}
\multiput(1092,178)(20.756,0.000){0}{\usebox{\plotpoint}}
\put(1110.00,178.00){\usebox{\plotpoint}}
\put(1130.75,178.00){\usebox{\plotpoint}}
\multiput(1131,178)(20.756,0.000){0}{\usebox{\plotpoint}}
\put(1151.51,178.00){\usebox{\plotpoint}}
\multiput(1156,178)(20.756,0.000){0}{\usebox{\plotpoint}}
\put(1172.26,178.00){\usebox{\plotpoint}}
\put(1193.02,178.00){\usebox{\plotpoint}}
\multiput(1194,178)(20.756,0.000){0}{\usebox{\plotpoint}}
\put(1213.77,178.00){\usebox{\plotpoint}}
\multiput(1220,178)(20.756,0.000){0}{\usebox{\plotpoint}}
\put(1234.53,178.00){\usebox{\plotpoint}}
\put(1255.29,178.00){\usebox{\plotpoint}}
\multiput(1258,178)(20.756,0.000){0}{\usebox{\plotpoint}}
\put(1276.04,178.00){\usebox{\plotpoint}}
\multiput(1283,178)(20.756,0.000){0}{\usebox{\plotpoint}}
\put(1296.80,178.00){\usebox{\plotpoint}}
\put(1317.55,178.00){\usebox{\plotpoint}}
\multiput(1321,178)(20.756,0.000){0}{\usebox{\plotpoint}}
\put(1338.31,178.00){\usebox{\plotpoint}}
\put(1359.06,178.00){\usebox{\plotpoint}}
\multiput(1360,178)(20.756,0.000){0}{\usebox{\plotpoint}}
\put(1379.82,178.00){\usebox{\plotpoint}}
\multiput(1385,178)(20.756,0.000){0}{\usebox{\plotpoint}}
\put(1400.57,178.00){\usebox{\plotpoint}}
\put(1421.33,178.00){\usebox{\plotpoint}}
\multiput(1423,178)(20.756,0.000){0}{\usebox{\plotpoint}}
\put(1436,178){\usebox{\plotpoint}}
\put(1306,812){\makebox(0,0)[r]{$V_m$}}
\put(1328.0,812.0){\rule[-0.200pt]{15.899pt}{0.400pt}}
\put(238,870){\usebox{\plotpoint}}
\multiput(238.58,861.60)(0.497,-2.424){59}{\rule{0.120pt}{2.023pt}}
\multiput(237.17,865.80)(31.000,-144.802){2}{\rule{0.400pt}{1.011pt}}
\multiput(269.58,714.77)(0.497,-1.763){61}{\rule{0.120pt}{1.500pt}}
\multiput(268.17,717.89)(32.000,-108.887){2}{\rule{0.400pt}{0.750pt}}
\multiput(301.58,604.14)(0.497,-1.347){59}{\rule{0.120pt}{1.171pt}}
\multiput(300.17,606.57)(31.000,-80.570){2}{\rule{0.400pt}{0.585pt}}
\multiput(332.58,522.42)(0.497,-0.957){61}{\rule{0.120pt}{0.863pt}}
\multiput(331.17,524.21)(32.000,-59.210){2}{\rule{0.400pt}{0.431pt}}
\multiput(364.58,462.12)(0.497,-0.743){59}{\rule{0.120pt}{0.694pt}}
\multiput(363.17,463.56)(31.000,-44.561){2}{\rule{0.400pt}{0.347pt}}
\multiput(395.58,416.77)(0.497,-0.546){61}{\rule{0.120pt}{0.538pt}}
\multiput(394.17,417.88)(32.000,-33.884){2}{\rule{0.400pt}{0.269pt}}
\multiput(427.00,382.92)(0.592,-0.497){51}{\rule{0.574pt}{0.120pt}}
\multiput(427.00,383.17)(30.808,-27.000){2}{\rule{0.287pt}{0.400pt}}
\multiput(459.00,355.92)(0.740,-0.496){39}{\rule{0.690pt}{0.119pt}}
\multiput(459.00,356.17)(29.567,-21.000){2}{\rule{0.345pt}{0.400pt}}
\multiput(490.00,334.92)(0.949,-0.495){31}{\rule{0.853pt}{0.119pt}}
\multiput(490.00,335.17)(30.230,-17.000){2}{\rule{0.426pt}{0.400pt}}
\multiput(522.00,317.92)(1.121,-0.494){25}{\rule{0.986pt}{0.119pt}}
\multiput(522.00,318.17)(28.954,-14.000){2}{\rule{0.493pt}{0.400pt}}
\multiput(553.00,303.92)(1.486,-0.492){19}{\rule{1.264pt}{0.118pt}}
\multiput(553.00,304.17)(29.377,-11.000){2}{\rule{0.632pt}{0.400pt}}
\multiput(585.00,292.93)(2.013,-0.488){13}{\rule{1.650pt}{0.117pt}}
\multiput(585.00,293.17)(27.575,-8.000){2}{\rule{0.825pt}{0.400pt}}
\multiput(616.00,284.93)(2.079,-0.488){13}{\rule{1.700pt}{0.117pt}}
\multiput(616.00,285.17)(28.472,-8.000){2}{\rule{0.850pt}{0.400pt}}
\multiput(648.00,276.93)(2.751,-0.482){9}{\rule{2.167pt}{0.116pt}}
\multiput(648.00,277.17)(26.503,-6.000){2}{\rule{1.083pt}{0.400pt}}
\multiput(679.00,270.93)(3.493,-0.477){7}{\rule{2.660pt}{0.115pt}}
\multiput(679.00,271.17)(26.479,-5.000){2}{\rule{1.330pt}{0.400pt}}
\multiput(711.00,265.93)(3.382,-0.477){7}{\rule{2.580pt}{0.115pt}}
\multiput(711.00,266.17)(25.645,-5.000){2}{\rule{1.290pt}{0.400pt}}
\multiput(742.00,260.94)(4.575,-0.468){5}{\rule{3.300pt}{0.113pt}}
\multiput(742.00,261.17)(25.151,-4.000){2}{\rule{1.650pt}{0.400pt}}
\multiput(774.00,256.95)(6.714,-0.447){3}{\rule{4.233pt}{0.108pt}}
\multiput(774.00,257.17)(22.214,-3.000){2}{\rule{2.117pt}{0.400pt}}
\multiput(805.00,253.95)(6.937,-0.447){3}{\rule{4.367pt}{0.108pt}}
\multiput(805.00,254.17)(22.937,-3.000){2}{\rule{2.183pt}{0.400pt}}
\multiput(837.00,250.95)(6.714,-0.447){3}{\rule{4.233pt}{0.108pt}}
\multiput(837.00,251.17)(22.214,-3.000){2}{\rule{2.117pt}{0.400pt}}
\put(868,247.17){\rule{6.500pt}{0.400pt}}
\multiput(868.00,248.17)(18.509,-2.000){2}{\rule{3.250pt}{0.400pt}}
\put(900,245.17){\rule{6.300pt}{0.400pt}}
\multiput(900.00,246.17)(17.924,-2.000){2}{\rule{3.150pt}{0.400pt}}
\put(931,243.67){\rule{7.709pt}{0.400pt}}
\multiput(931.00,244.17)(16.000,-1.000){2}{\rule{3.854pt}{0.400pt}}
\put(963,242.17){\rule{6.500pt}{0.400pt}}
\multiput(963.00,243.17)(18.509,-2.000){2}{\rule{3.250pt}{0.400pt}}
\put(995,240.67){\rule{7.468pt}{0.400pt}}
\multiput(995.00,241.17)(15.500,-1.000){2}{\rule{3.734pt}{0.400pt}}
\put(1026,239.17){\rule{6.500pt}{0.400pt}}
\multiput(1026.00,240.17)(18.509,-2.000){2}{\rule{3.250pt}{0.400pt}}
\put(1058,237.67){\rule{7.468pt}{0.400pt}}
\multiput(1058.00,238.17)(15.500,-1.000){2}{\rule{3.734pt}{0.400pt}}
\put(1089,236.67){\rule{7.709pt}{0.400pt}}
\multiput(1089.00,237.17)(16.000,-1.000){2}{\rule{3.854pt}{0.400pt}}
\put(1121,235.67){\rule{7.468pt}{0.400pt}}
\multiput(1121.00,236.17)(15.500,-1.000){2}{\rule{3.734pt}{0.400pt}}
\put(1152,234.67){\rule{7.709pt}{0.400pt}}
\multiput(1152.00,235.17)(16.000,-1.000){2}{\rule{3.854pt}{0.400pt}}
\put(1215,233.67){\rule{7.709pt}{0.400pt}}
\multiput(1215.00,234.17)(16.000,-1.000){2}{\rule{3.854pt}{0.400pt}}
\put(1247,232.67){\rule{7.468pt}{0.400pt}}
\multiput(1247.00,233.17)(15.500,-1.000){2}{\rule{3.734pt}{0.400pt}}
\put(1184.0,235.0){\rule[-0.200pt]{7.468pt}{0.400pt}}
\put(1310,231.67){\rule{7.468pt}{0.400pt}}
\multiput(1310.00,232.17)(15.500,-1.000){2}{\rule{3.734pt}{0.400pt}}
\put(1278.0,233.0){\rule[-0.200pt]{7.709pt}{0.400pt}}
\put(1373,230.67){\rule{7.468pt}{0.400pt}}
\multiput(1373.00,231.17)(15.500,-1.000){2}{\rule{3.734pt}{0.400pt}}
\put(1341.0,232.0){\rule[-0.200pt]{7.709pt}{0.400pt}}
\put(1404.0,231.0){\rule[-0.200pt]{7.709pt}{0.400pt}}
\end{picture}

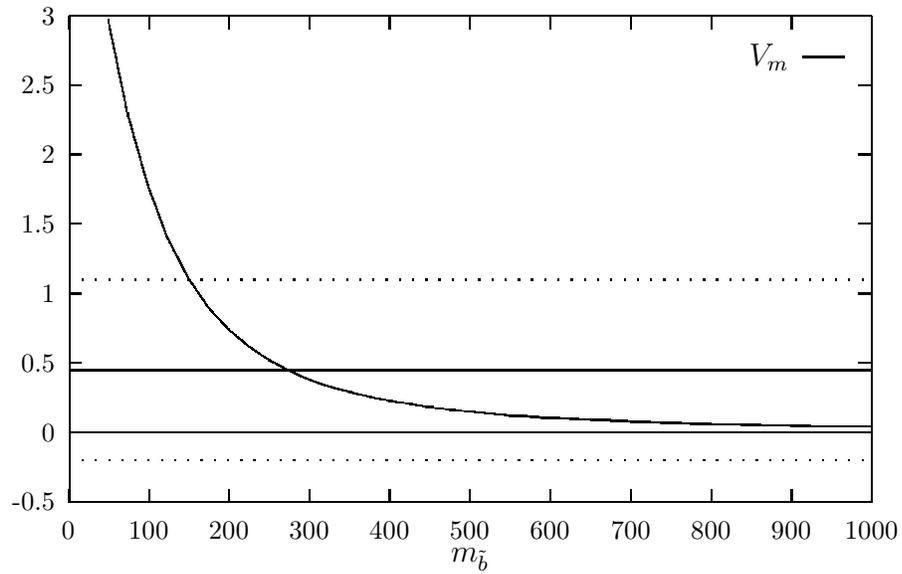
Figure 3: The same as Figure 1 for $V_M$. $V_m^{exp}-V_m^{theor}=0.45(65)$ was used.
\vspace{8mm}
\newpage

Now let us take into account $\tilde{t}_L \tilde{t}_R$ mixing which
in general is not small being of the order of $t$-quark mass times
the superpartner mass scale, $m^2_{LR} \sim m_t m_{SUSY}$. As a
result of mixing two states with masses $m_1$ and $m_2$ are formed,
$\tilde{t}_1 = c_u \tilde{t}_L + s_u \tilde{t}_R$ and
$\tilde{t}_2 = -s_u \tilde{t}_L + c_u \tilde{t}_R$,
where $c_u \equiv \cos\theta_{LR}$, $s_u \equiv \sin\theta_{LR}$ and
our functions $\delta_{SUSY}^{LR}V_i$ depend on 4 parameters: $m_1$,
$m_2$, $\theta_{LR}$ and $m_{\tilde{b}}$, three of which are
independent (the difference of $m^2_{\tilde{t}_L \tilde{t}_L}$ and
$m^2_{\tilde{b}_L \tilde{b}_L}$ is determined by the top quark mass).
The numerical values of these parameters depend on the SUSY model.
Let us present the formulas for the electroweak radiative corrections
which take $\tilde{t}_L \tilde{t}_R$ mixing into account:
\begin{equation}
\delta_{SUSY}^{LR} V_A = \frac{1}{m_Z^2} [c_u^2 g(m_1, m_{\tilde{b}})
+ s_u^2 g(m_2, m_{\tilde{b}}) - c_u^2 s_u^2 g(m_1, m_2)] \;\; ,
\label{9}
\end{equation}

\begin{equation}
\delta_{SUSY}^{LR} V_R = \delta_{SUSY}^{LR}V_A + \frac{1}{3} Y_L
[c_u^2 \ln(\frac{m_1^2}{m^2_{\tilde{b}}})
+ s_u^2 \ln(\frac{m_2^2}{m^2_{\tilde{b}}})] - \frac{1}{3} c_u^2 s_u^2
h(m_1, m_2) \;\; ,
\label{10}
\end{equation}

\begin{eqnarray}
\delta_{SUSY}^{LR} V_m &=& \delta_{SUSY}^{LR}V_A + \frac{2}{3} Y_L
s^2[c_u^2 \ln(\frac{m_1^2}{m^2_{\tilde{b}}})
+ s_u^2 \ln(\frac{m_2^2}{m^2_{\tilde{b}}})] + \nonumber \\
&+& \frac{c^2 -s^2}{3}
[c_u^2 h(m_1, m_{\tilde{b}}) +  
 s_u^2 h(m_2, m_{\tilde{b}})] - \frac{c_u^2 s_u^2}{3} h(m_1, m_2)
\;\; .
\label{11}
\end{eqnarray}

Our approximation should be  good for the case when all superpartners 
have more or less equal masses. When some spartners are considerably
lighter than sbottom their contribution can dominate and our 
approximation can fails.

It is desirable to compare numerically the formulas for the enhanced
radiative corrections with the results of full one loop calculations
in SUSY  models in order to check how good our approximation is.

We are grateful to Z.Berezhiani for discussions on SUSY models.
Investigations of I.G., A.N., V.N. and M.V. were supported by RFBR
grants 96-02-18010 and 96-15-96578; those of A.N., V.N. and M.V. by
INTAS grants 93-3316 and 94-2352 and INTAS-RFBR grant 95-05678 as well. 
V.A.Novikov
is grateful to G.Fiorentini and L.Piemontese for hospitality in Ferrara,
where this work was finished.

\end{document}